\documentclass[twoside,slac_one]{revtex4}
\usepackage{graphicx}
\usepackage{fancyhdr}
\usepackage{amsmath} 
\usepackage{bm}
\usepackage{amsxtra}
\usepackage{amssymb}
\usepackage{amsthm}
\usepackage{latexsym}
\usepackage{lscape}
\usepackage{xspace}

\newcommand{\qqbar}{\ensuremath{q\bar{q}}\xspace}
\newcommand{\ttbar}{\ensuremath{t\bar{t}}\xspace}
\newcommand{\invfb}{\ensuremath{\mathrm{fb}^{-1}}\xspace}
\newcommand{\invpb}{\ensuremath{\mathrm{pb}^{-1}}\xspace}

\begin{document}

\title{Top Quark Results Using CMS Data at 7 TeV}

%

\author{K. M. Ecklund}
\affiliation{Department of Physics and Astronomy, Rice University, Houston, TX, USA}

\begin{abstract}
I give an overview of recent results on top quark properties
and interactions, obtained using data collected with the CMS
experiment during the years 2010--2011 at $\sqrt{s}=$ 7
TeV. Measurements are presented for the inclusive top pair 
production cross section, using the dilepton, lepton plus jets, and hadronic
channels. The mass of the top quark is measured using the
dilepton and lepton plus jets samples.   
CMS also measures the cross section for electroweak production
of single top quarks and constrains the CKM
matrix element $V_{tb}$.  Top quark results are compared with Standard
Model predictions and used to search for possible presence of new
physics. In particular, measurements of the top-pair invariant mass
distribution are used to search for new particles decaying to top
pairs. CMS has also investigated the top-pair charge asymmetry to
search for possible new physics contributions.

\end{abstract}

\maketitle

\thispagestyle{fancy}


\section{Introduction}

As the heaviest known quark, top presents an opportunity for Large
Hadron Collider (LHC) experiments to study the electroweak symmetry
breaking sector of the Standard Model (SM) and to search for physics
beyond the SM.  The LHC is a top factory even at $\sqrt{s}=7$ TeV,
with more than 10$^5$ \ttbar pairs produced per \invfb.  Statistics of
top samples at the LHC already match or exceed samples produced at the
Tevatron, and in time the systematic uncertainties will rival those
from the well established CDF and D0 experiments.  The samples
collected to date by the Compact Muon Solenoid experiment (CMS) \cite{cms}
include 36 \invpb from the 2010 7 TeV run and already more than 1
\invfb integrated in the ongoing 2011 run, which has a nominal goal of 5 \invfb.
Most results presented here are based on the 2010 data set, but a few
analyses have included the available 2011 data already.

With these samples, the SM is tested in both top production
and top decay.  Both production of \ttbar pairs and electroweak
production of single top quarks at $\sqrt{s}=7$ TeV have been measured
and compared with SM predictions.  Top quarks may also
appear in decays of new particles, e.g. heavy $Z^\prime$ or fourth
generation quarks.  New physics could also modify top quark couplings,
so large samples provide an opportunity to limit new physics
contributions in top production and decay by examining asymmetries and
differential distributions.

The relation between the top quark mass and radiative corrections to
the mass of a SM Higgs boson is well known.  As large samples of top
quarks are produced at the LHC, mass measurements can constrain the SM
Higgs mass before a discovery and provide a test of the internal
consistency of the electroweak model after any Higgs boson discovery.
Present top mass measurements from CMS are based on established methods from
the Tevatron and show the potential of the detector.

\section{Top Quark Production}
Top quark production in the SM is dominated by the QCD quark
annihilation ($\qqbar\to\ttbar$) and gluon fusion ($gg\to\ttbar$)
processes.  At the LHC, the $pp$ initial state gives parton-parton
luminosities that favor gluon fusion ($\sim 85$\%) over quark
annihilation, providing complementarity to the Tevatron where the
situation is reversed.  Standard Model cross sections are computed by
several groups, including some of the next-to-next-to-leading order
contributions \cite{xstTtheory1,xstTtheory2,xstTtheory3}, giving
$\sigma_{\ttbar} = 163^{+11}_{-10}$ pb for $pp$ collisions at 7 TeV. 

Single top quarks are also produced by electroweak processes in the
$t$ channel, from intrinsic $b$ (sea quark) content in the proton or
via gluon splitting, and via the $s$-channel $W$ exchange.  Cross
sections here are smaller, with the $t$ channel dominating at 62 pb.

With $V_{tb}\approx 1$, top quarks decay to $Wb$ 100\% of the time.  Top
pair decays are characterized by the leptonic or hadronic decay of the
two $W$s, for topologies of dilepton, lepton+jets, and fully hadronic.
The leptons may be electron, muon, or tau and are accompanied by a
neutrino that may be identified experimentally by missing transverse energy.

\subsection{Top pair cross section}
\begin{figure}[ht]
\centering
\includegraphics[width=0.48\textwidth]{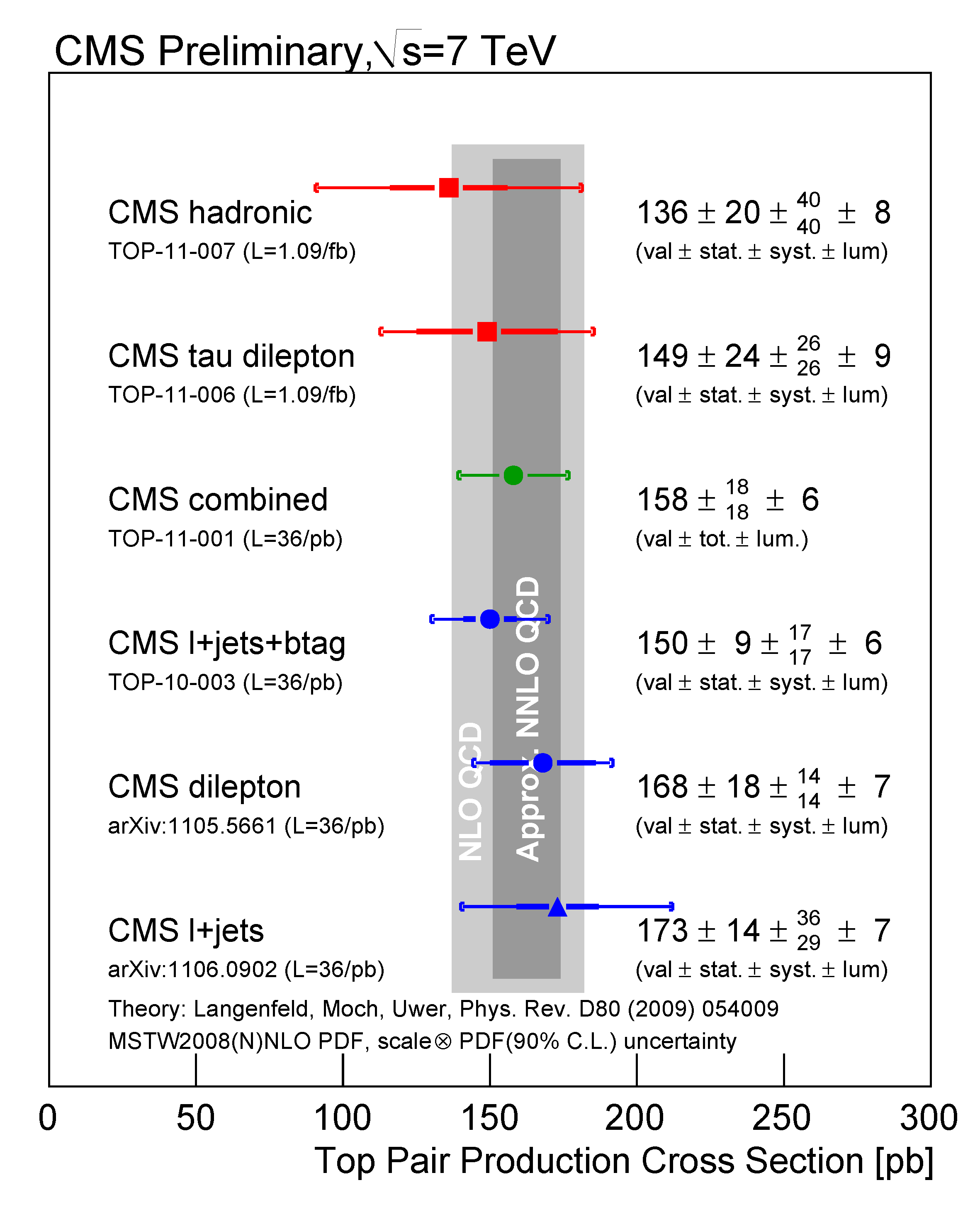}
\includegraphics[width=0.48\textwidth]{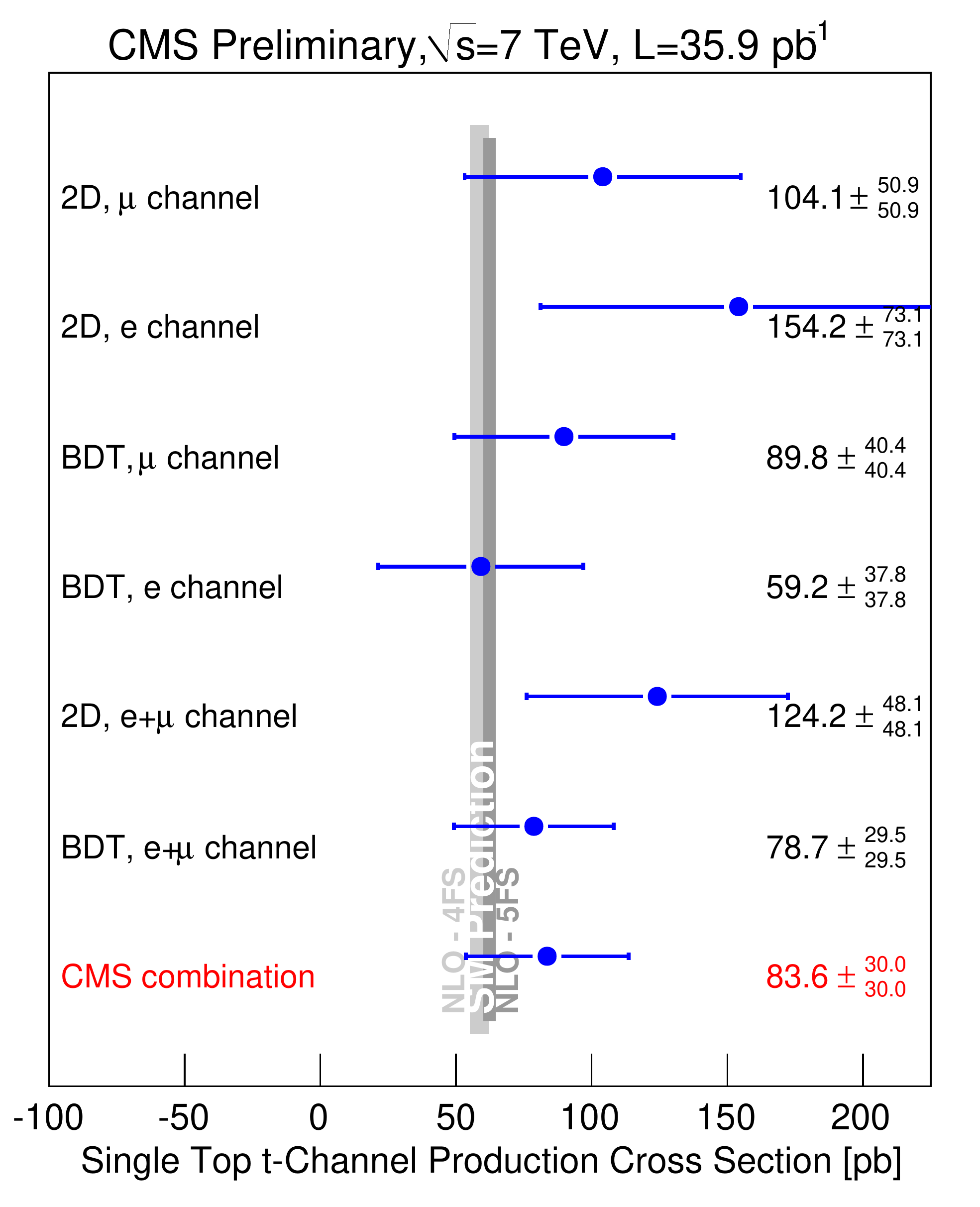}
\caption{Summary of \ttbar (left) and single top (right) production cross section measurements from CMS.} \label{fig:top_xs}
\end{figure}

Using the 2010 data sample (36 \invpb), CMS measured \ttbar production
using the dilepton \cite{top-11-002} channel and the lepton+jets channel 
with \cite{top-10-003} and without \cite{top-10-002} b-jet identification.  
New for 2011 are a cross section measurement using the fully hadronic
channel \cite{top-11-007} and one with dileptons using one identified
tau lepton and a muon \cite{top-11-006}.  The production cross sections
measured in all channels are in good agreement with one another and
SM calculations.
Figure~\ref{fig:top_xs} displays the CMS measurements in each
channel and the CMS combined result of
$\sigma_{\ttbar}=158\pm18(\textrm{stat.+syst.})\pm6(\textrm{lumi.})$ pb
from the 2010 data, overlaid with the NLO and (N)NLO calculations.
Details of the \ttbar cross section measurement were presented at this
conference by S.~Khalil \cite{khalil}.

\subsection{Single top cross section}
CMS has also measured single top $t$-channel production in $pp$
collisions at 7~TeV \cite{singletop}.  The $t$ channel has the largest
expected cross section (62 pb) and 
a distinctive kinematic signature of a forward light-flavor jet from
the $t$-channel exchange of a virtual $W$, making this a favorable
channel.  Two approaches are used to isolate the expected $t$-channel
signal from backgrounds in the decay chain $t\to W b \to \ell\nu b$.
One makes a two-dimensional fit to the pseudo-rapidity of the light
jet accompanying the top and the angle between the lepton from the $W$
and the jet, which has a characteristic shape from the $V-A$ current.
The other uses a multivariate technique (boosted decision tree) to
fully exploit the expected kinematics of single top production and top
decay.  Further details of these analyses were presented by T.~Speer
at this conference \cite{Speer}.

Both analyses find evidence for single top production, with a combined
measurement of $\sigma_t=83.6\pm30 \textrm{(stat.+syst.)}\pm
3(\textrm{lumi.})$ pb ($t$ channel only) for a combined significance of 3.5 standard
deviations.  Figure~\ref{fig:top_xs} shows the measurements from
the two analyses and the combination compared to SM calculations,
which are in good agreement for the large experimental uncertainties.
The comparison to the SM predicted cross section limits
$|V_{tb}| > 0.62$ at 95\% confidence level.  This is a direct
measurement of a tree-level process sensitive to $V_{tb}$, fully
consistent with indirect measurements from loop processes.

\section{Top Quark Mass Measurements}
\begin{figure}[ht]
\centering
\includegraphics[width=0.48\textwidth]{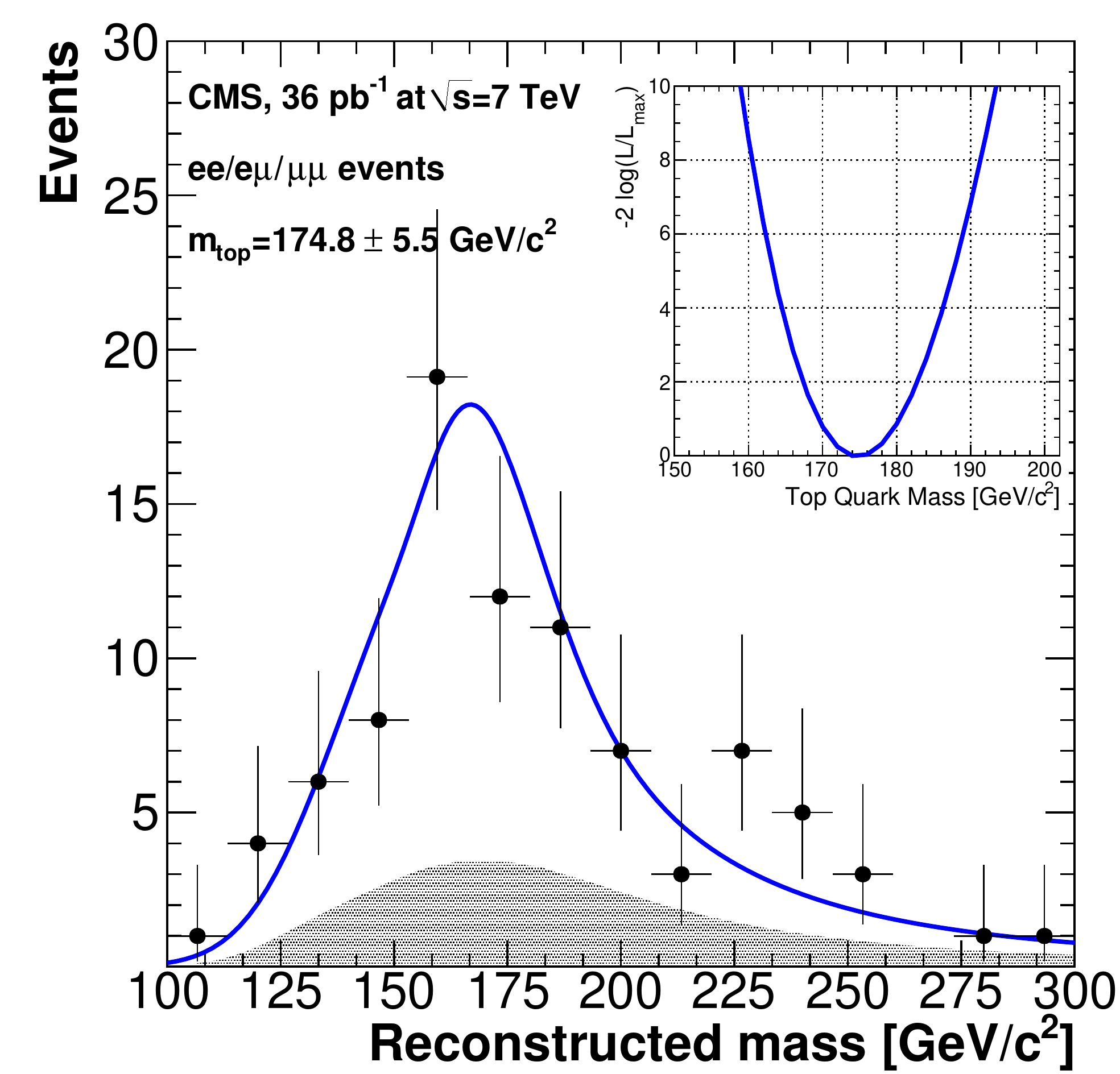}
\includegraphics[width=0.48\textwidth]{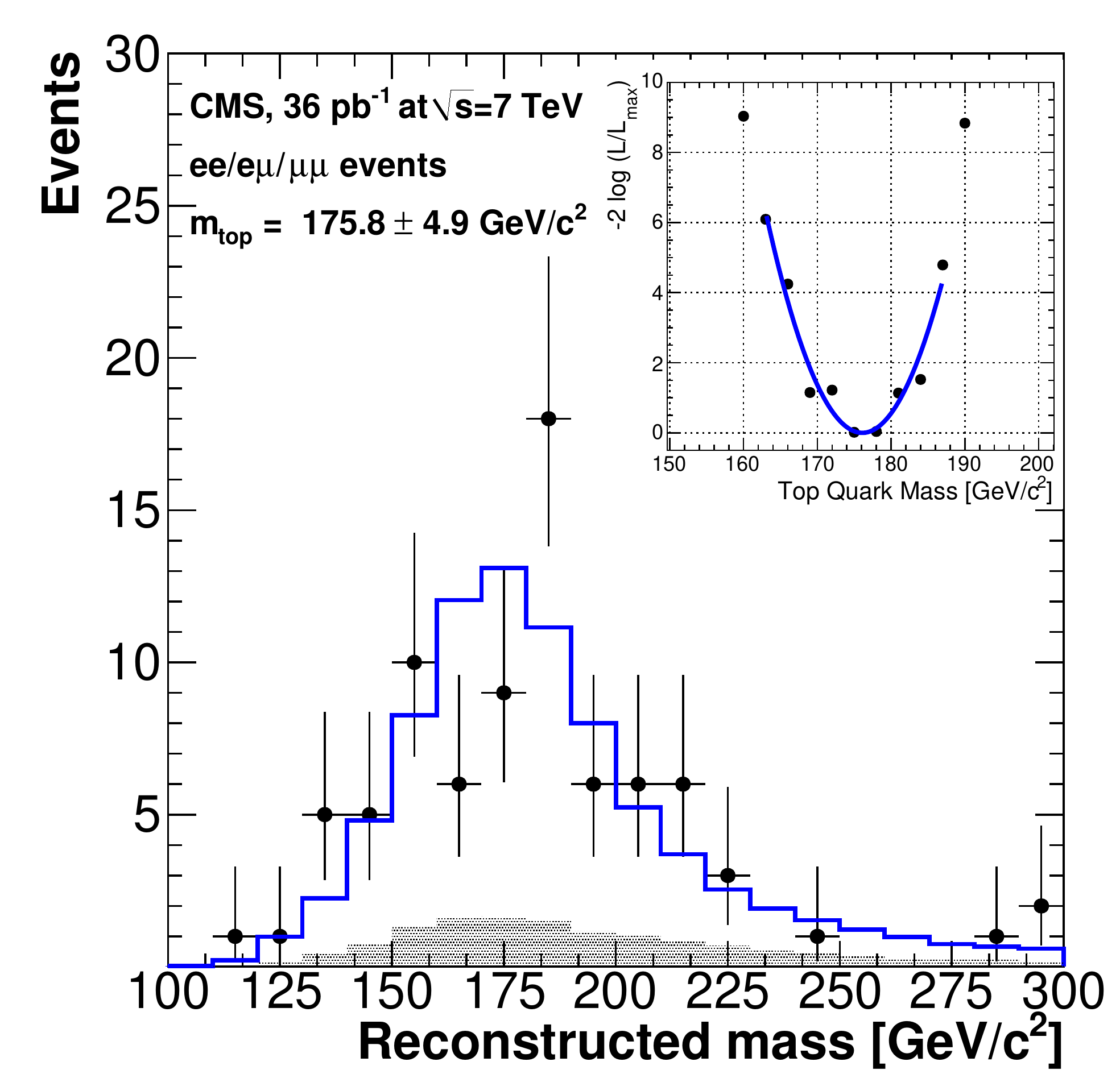}
\caption{Reconstructed top quark mass from kinematic fitting method
  (left) and matrix element weighting technique (right).} \label{fig:mtop_dilepton}
\end{figure}

CMS has used clean top quark samples from the dilepton and lepton plus
jets channels to measure the mass of the top quark.  In the dilepton
channel \cite{top-11-002}, 36 \invpb of data are analyzed using both a
kinematic fitting method and a matrix element weighting technique.
These techniques adapted from Tevatron analyses work well also at the
LHC.  Figure~\ref{fig:mtop_dilepton} shows the reconstructed top quark
mass using the two techniques.
The results are in good agreement with each other and the world
average.  When combined taking into account all correlations, the
top quark mass is found to be 
$m_t=175.5\pm4.6 (\textrm {stat.}) \pm 4.6 (\textrm {syst.})$ GeV. The
dominant systematic uncertainties come from the jet energy scales for
light and b-flavored jets, which are improvable in larger data
samples.

\begin{figure}[ht]
\centering
\includegraphics[height=0.48\textwidth,angle=90]{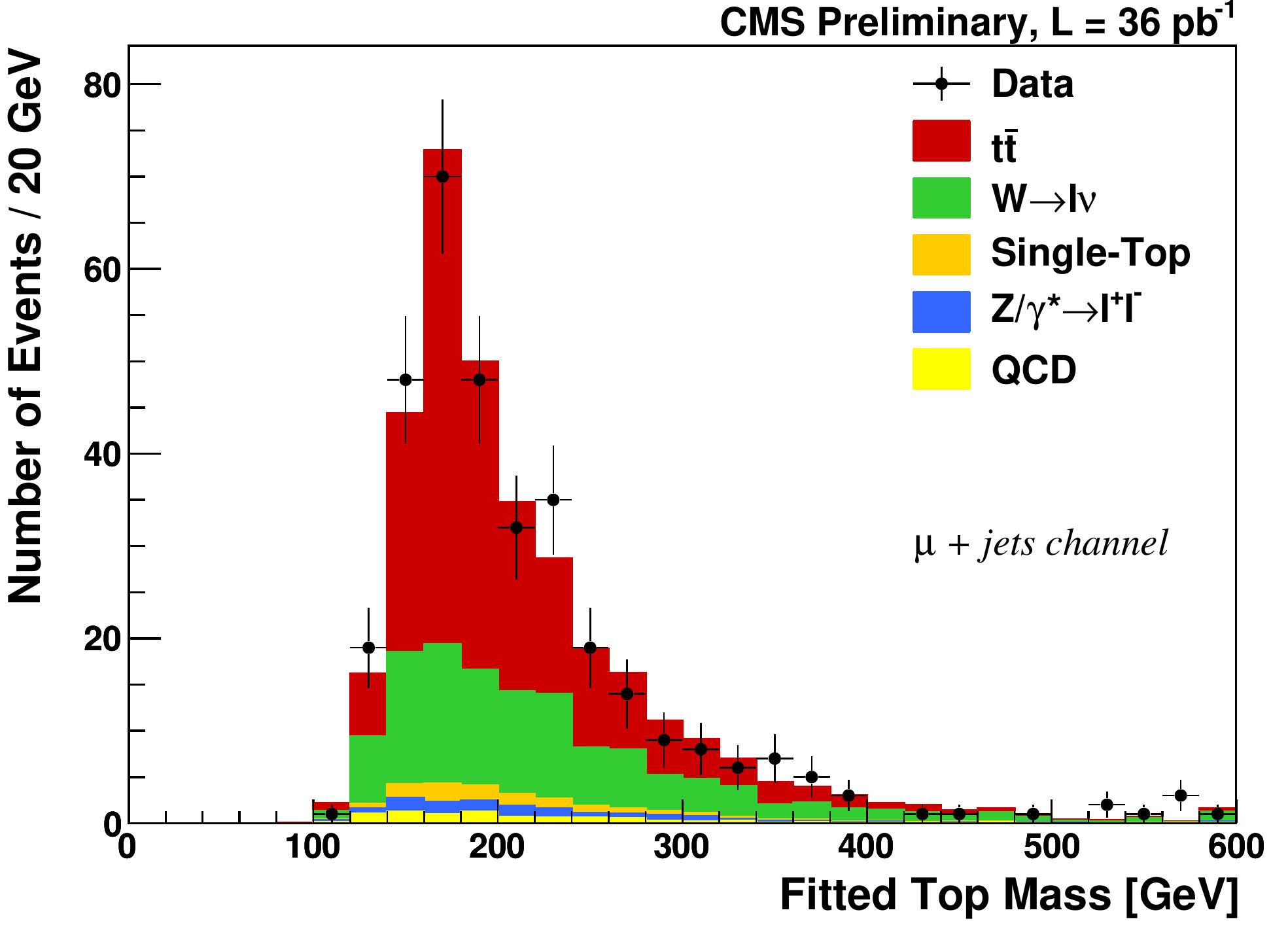}
\includegraphics[height=0.48\textwidth,angle=90]{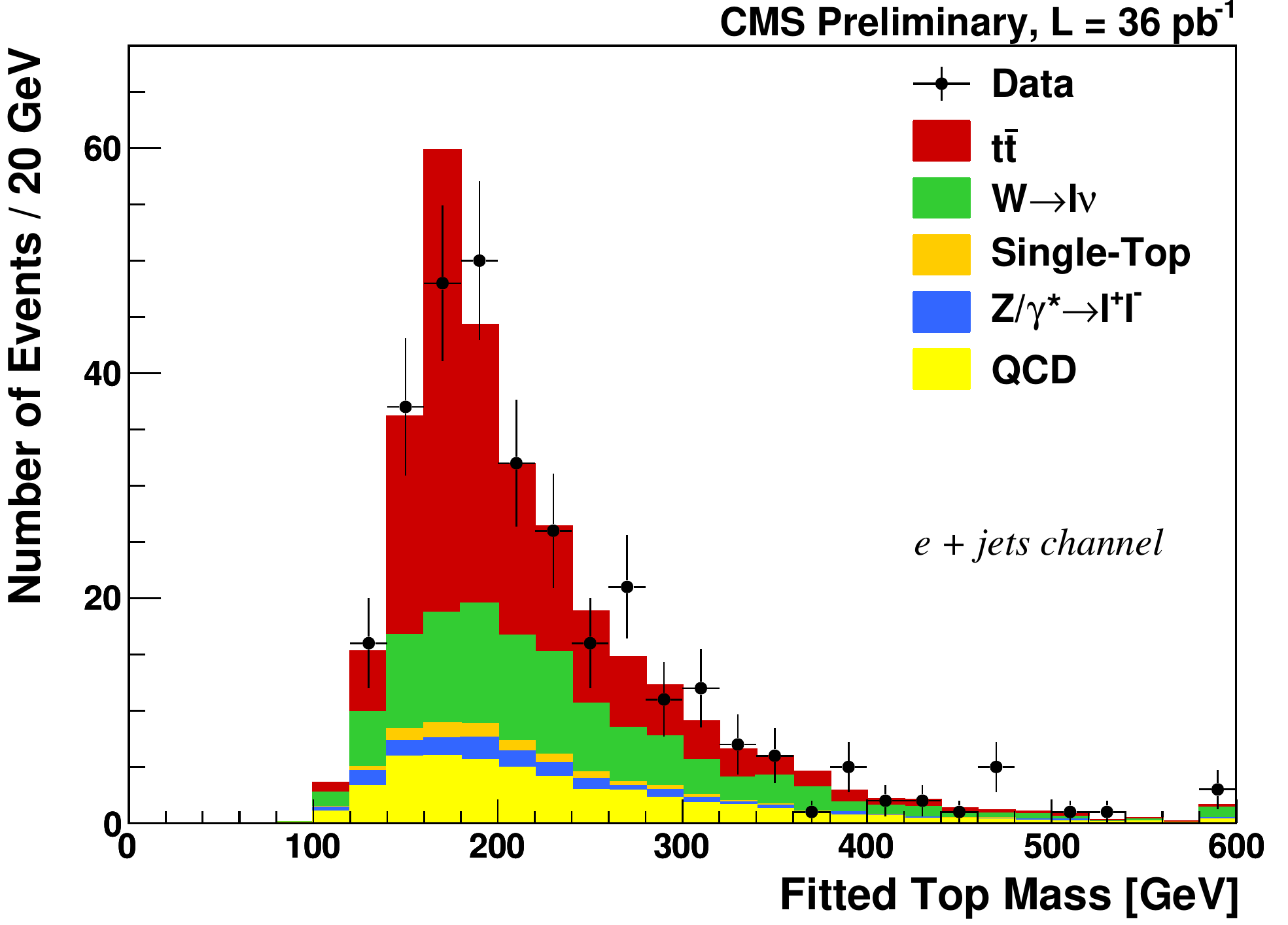}
\caption{Reconstructed top quark mass after the kinematic fit for
  muon+jets (left) and electron+jets (right) samples.} \label{fig:mtop_ideogram}
\end{figure}

Similarly, in the lepton plus jets channel, CMS has used two analysis
techniques to reconstruct the top quark mass from a kinematic fit to
the decay products \cite{top-10-009}.  
Using an ideogram method that combines the event-by-event likelihood for
an assumed top mass given the results of the kinematic fit, the most
precise result from the 36 \invpb sample is obtained: 
$m_t=173.1\pm2.1 (\textrm{stat.})^{+2.8}_{-2.5}(\textrm{syst.})$ GeV, 
using muons and electrons (See Fig.\ref{fig:mtop_ideogram}).  The CMS
combination with the dilepton result is 
$m_t=173.4\pm1.9 (\textrm{stat.}) \pm 2.7 (\textrm{syst.})$ GeV.
As a cross check, a template analysis jointly fits the jet-energy
scale and the top quark mass in the muon plus jets channel, with the
aim of reducing the leading systematic uncertainty by calibrating {\em
  in situ} with the $W$ mass constraint.  Results are consistent with
the ideogram method.

The CMS top mass results were covered in more detail in a dedicated
presentation at this meeting \cite{Avetisyan}.

\section{Top-Pair Invariant Mass Distribution}
\begin{figure}[ht]
\centering
\includegraphics[width=0.50\textwidth]{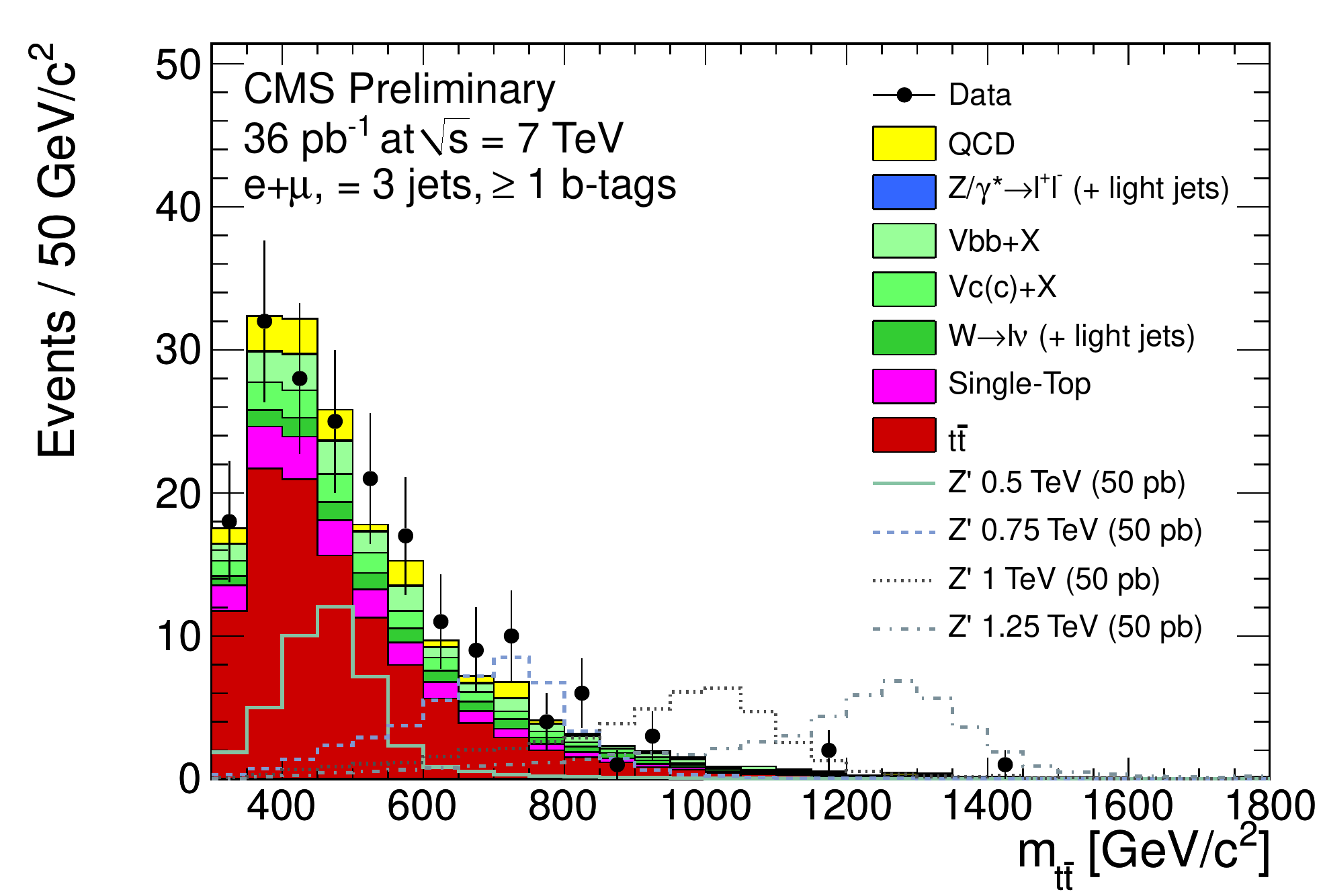}
\hfil
\includegraphics[width=0.37 \textwidth]{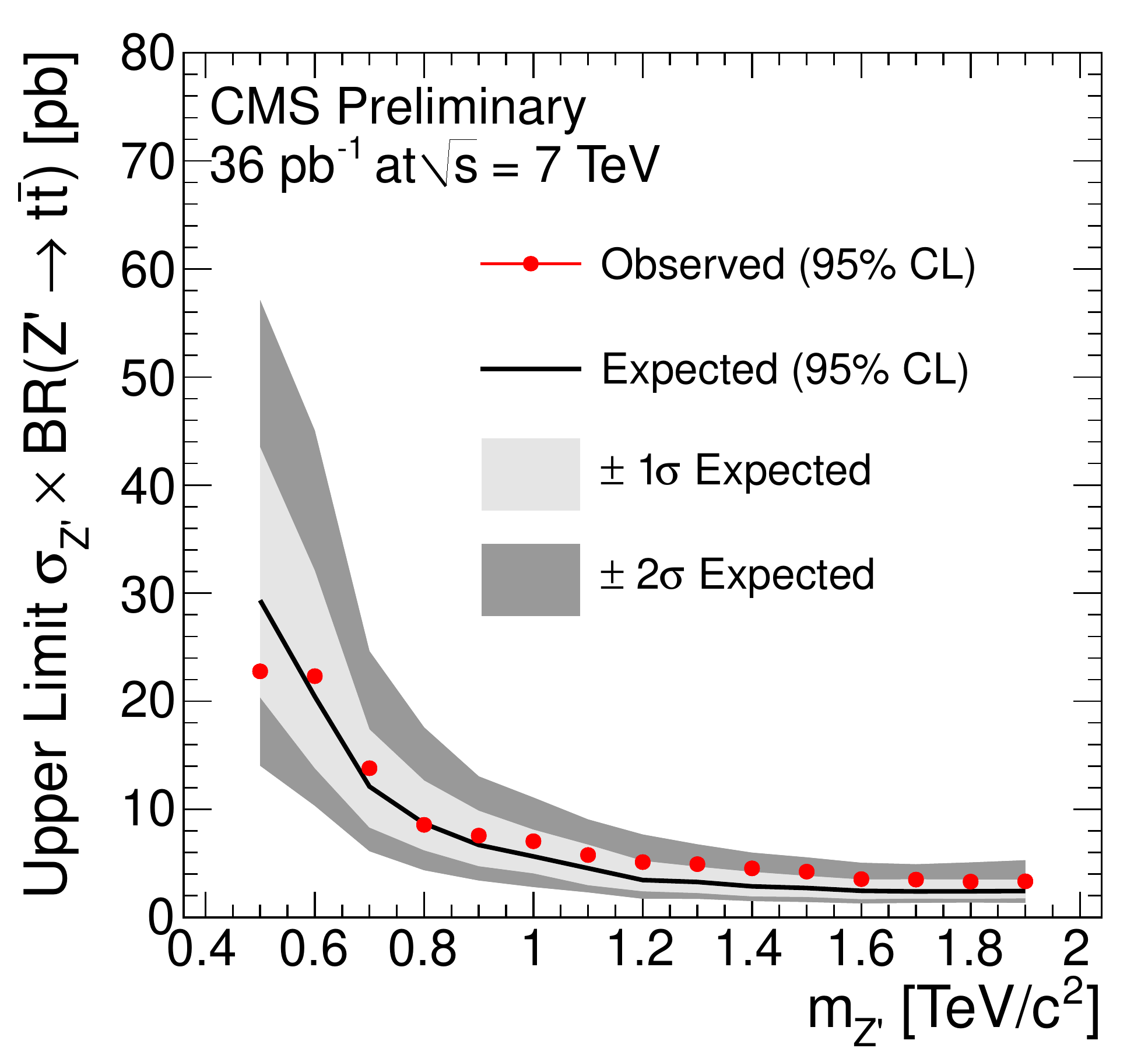}
\caption{Reconstructed \ttbar invariant mass in a lepton plus jet
  sample (left) and resulting limits on the production cross section
  $\times$ branching fraction for a narrow $Z^\prime$ (right).} \label{fig:mttbar}
\end{figure}

In \ttbar events, the invariant mass of the \ttbar pair may be
examined for resonances or other contributions from new physics
processes.  CMS has measured the top pair invariant mass and set
limits on production of narrow resonances such as a $Z^\prime$ or
Kaluza-Klein gluon.  From the 36 \invpb sample, the $m(\ttbar)$
spectrum in Fig.~\ref{fig:mttbar} is obtained by selecting
electron/muon plus jets events compatible with \ttbar production, with
no indications of any resonance \cite{top-10-007}.

\begin{figure}[htb]
\centering
\includegraphics[width=0.48\textwidth]{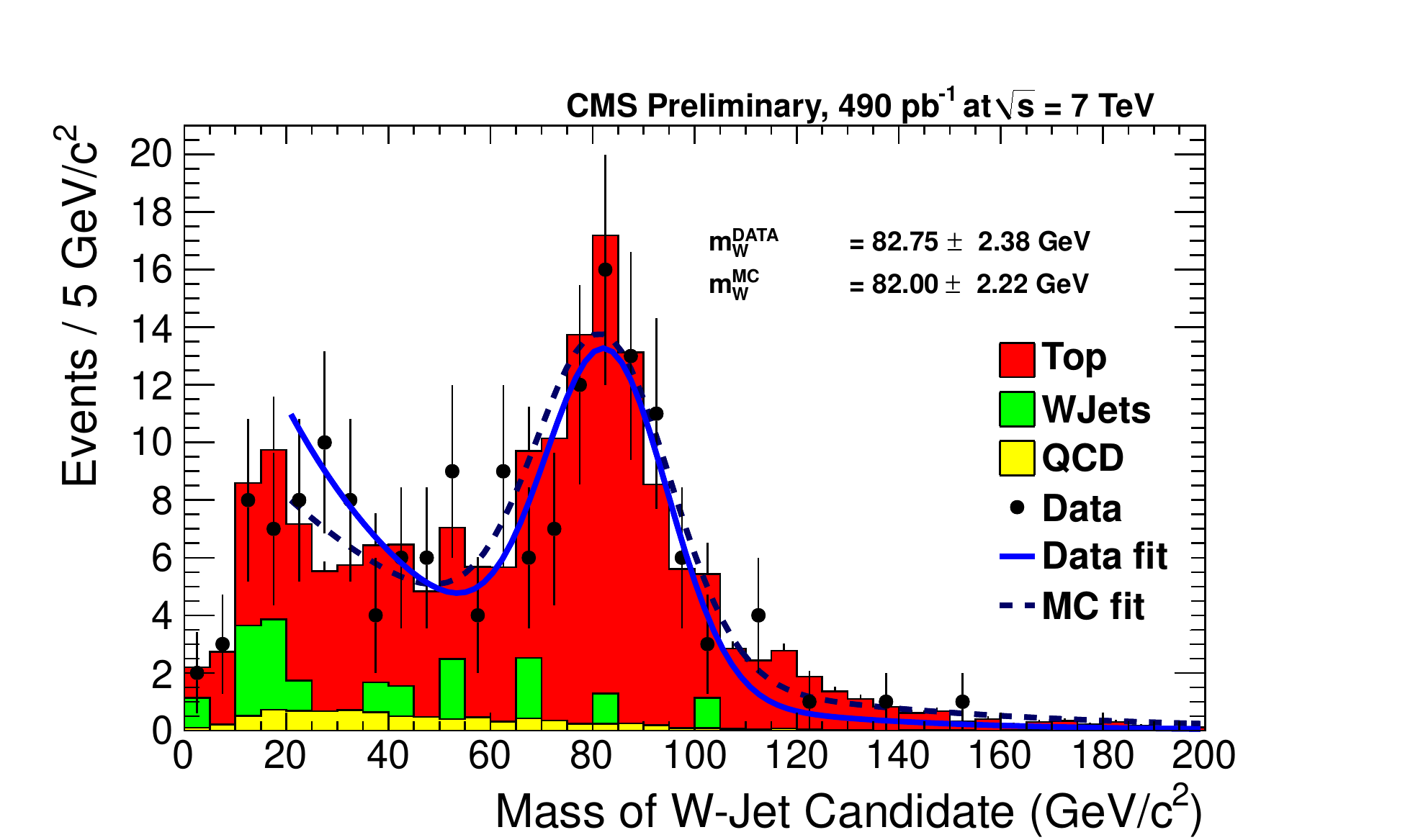}
\includegraphics[width=0.48\textwidth]{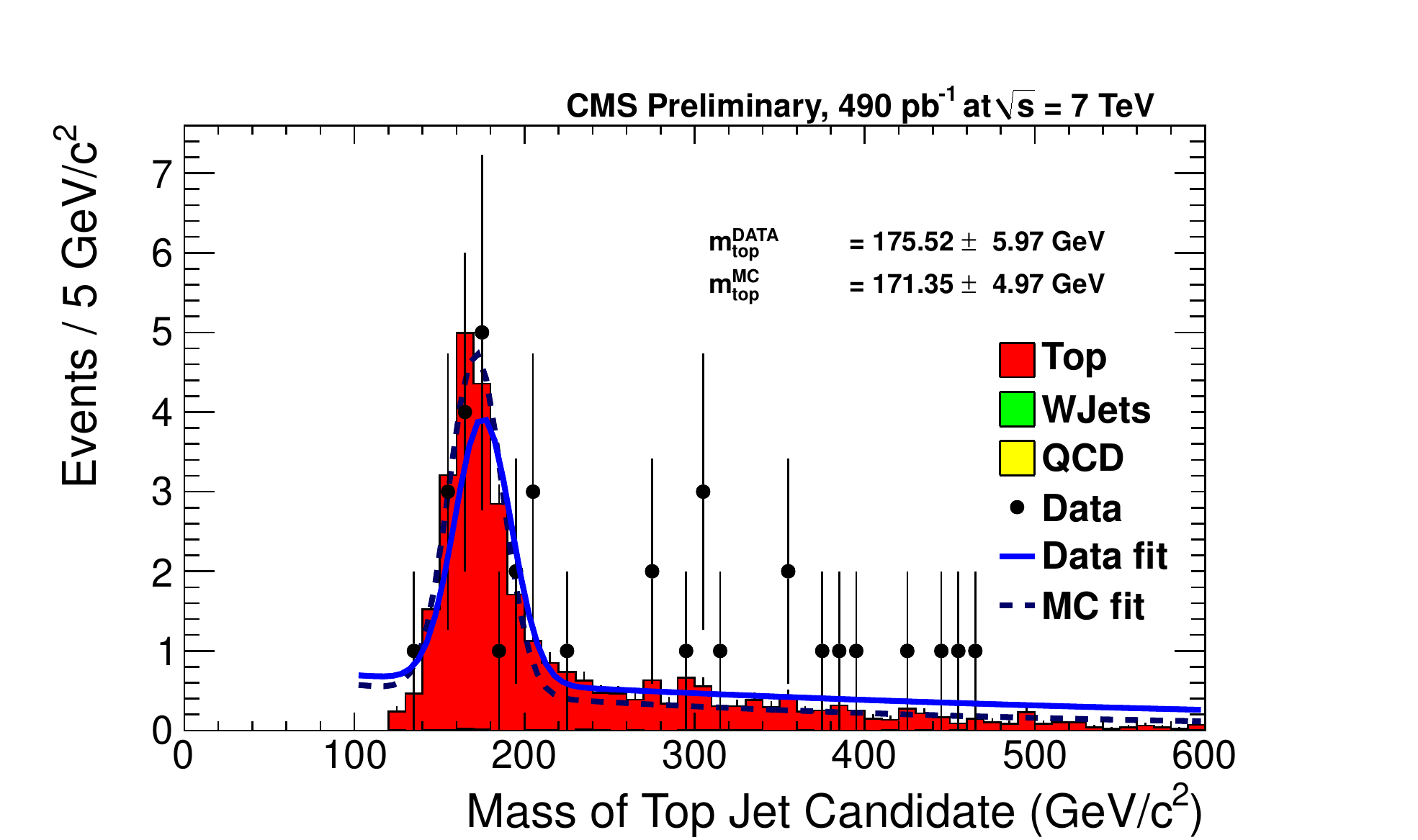}
\caption{Reconstruction of $W$ sub-jet (left) and merged top quark jet
  (right) in a muon+jets \ttbar sample.}
 \label{fig:top-jets}
\end{figure}

\begin{figure}[ht]
\centering
\includegraphics[width=0.48\textwidth]{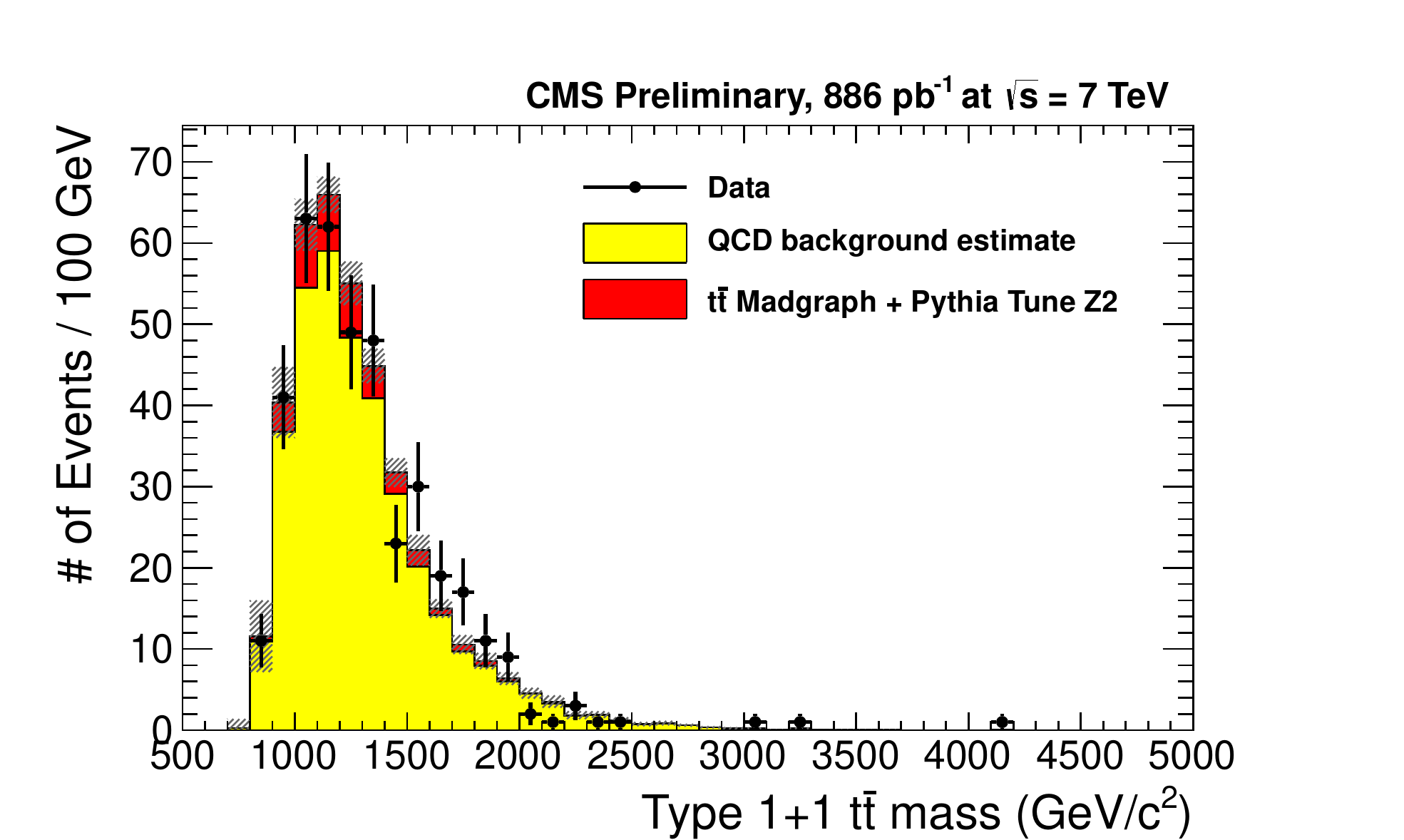}
\includegraphics[width=0.48\textwidth]{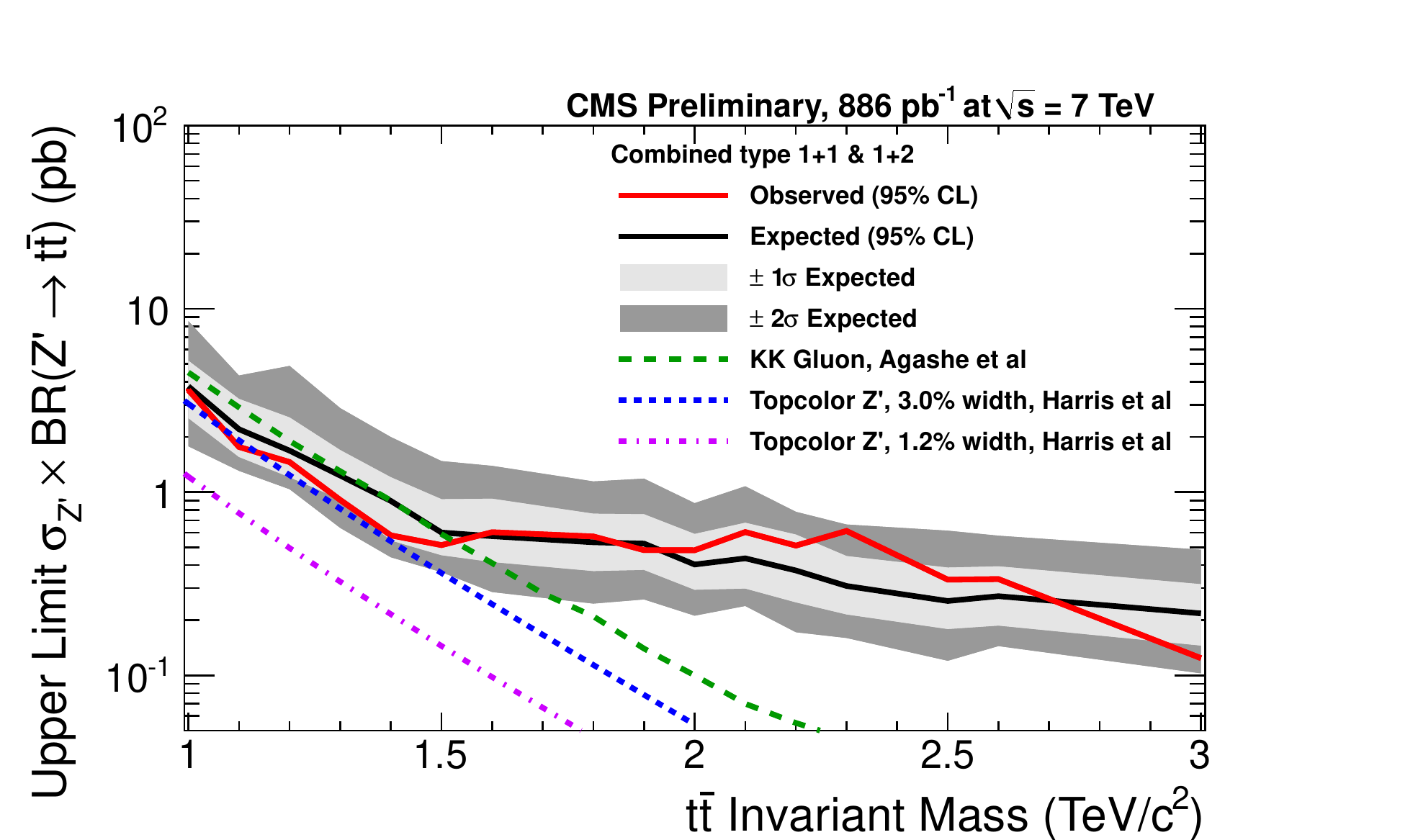}
\caption{ Invariant mass m(\ttbar) of hadronic \ttbar candidates (left) and limts on
  production of heavy resonances (right).
} \label{fig:boostedtop}
\end{figure}

CMS has also measured the \ttbar invariant mass spectrum in fully
hadronic events using a novel analysis that reconstructs highly
boosted top jets using jet substructure.  The hadronic $W$ decay is
resolved from sub-jets contained in the merged top jet.  The technique
is validated in a muon plus jets \ttbar sample
(Fig.~\ref{fig:top-jets}) and then applied to fully hadronic events
(Fig.~\ref{fig:boostedtop}) 
matching the boosted topology.  The
dominant QCD multijet background is estimated from the data.
Use of the boosted top reconstruction extends the sensitivity to higher mass
resonances.  See the presentation by S. Rappoccio at this meeting
\cite{Rappoccio} for additional details.


\section{Top Charge Asymmetry}
New physics can also appear as charge asymmetries in the top
(anti)quark rapidity distribution.  CDF \cite{cdf_Actop}
and D0 \cite{d0_Actop} have measured an
unexpectedly large forward-backward charge asymmetry in the top/antitop
rapidity distribution.  In $p\bar{p}$ collisions at the Tevatron, the SM
expectation is for a small asymmetry ($\sim 5$\%) due to interference
between leading order and box diagrams, and from interference of
initial- and final-state radiative diagrams, favoring top
quarks in the proton direction \cite{ac_SM}.  CDF reported a
larger asymmetry for large \ttbar invariant mass, suggesting influence of a heavy
particle.  For $pp$ collisions, the production via gluon fusion is symmetric,
though the non-dominant \qqbar annihilation would lead in this case to
$\sim 1$\% asymmetry \cite{ac_SM}, which is manifested in the width of the rapidity
distribution, since the initial state is symmetric.  If due to new
physics, e.g. additional interference contributions from axigluons,
LHC experiments may also observe an unexpectedly large asymmetry.

\begin{figure}[ht]
\centering
\includegraphics[width=0.48\textwidth]{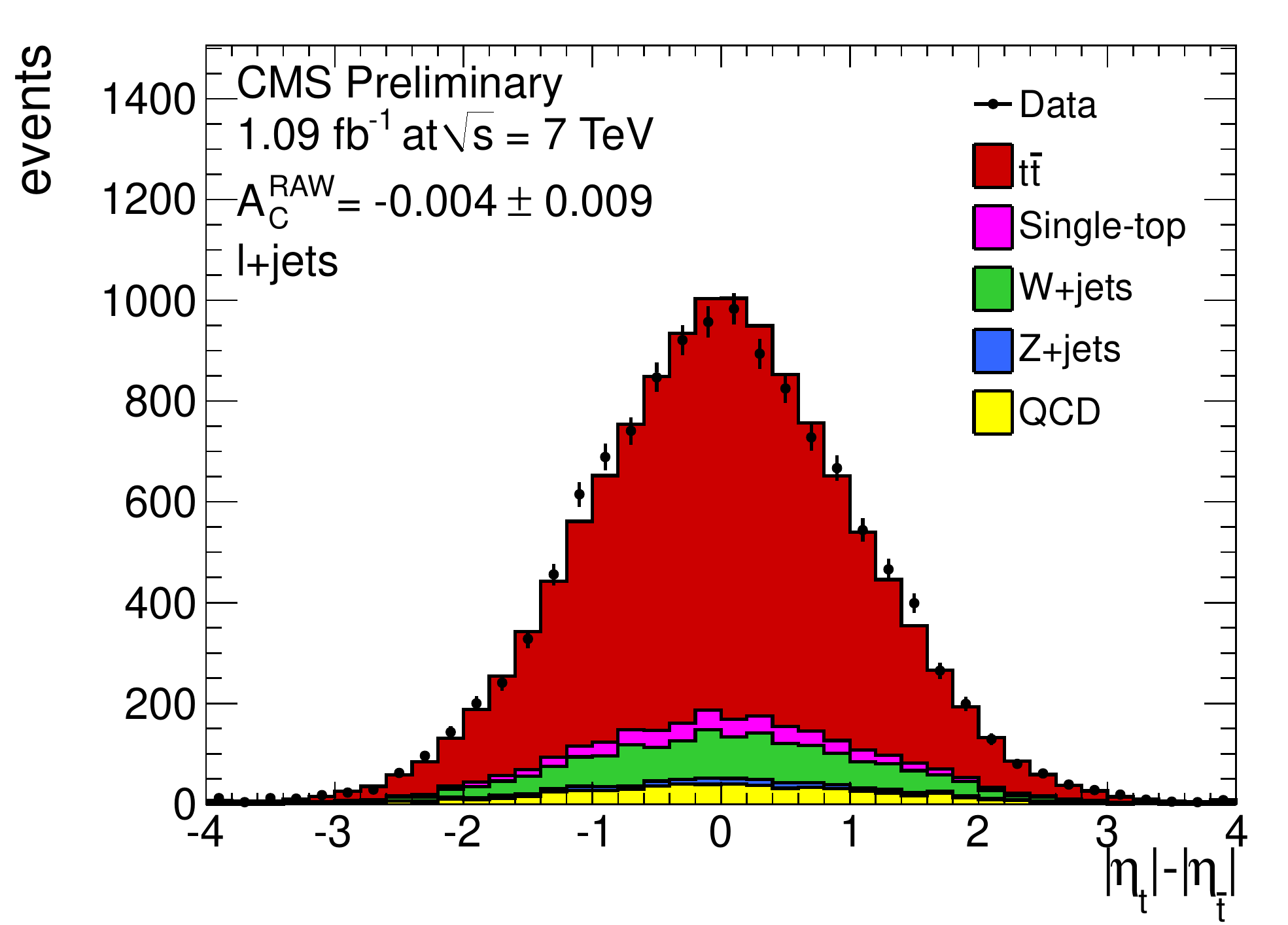}
\includegraphics[width=0.48\textwidth]{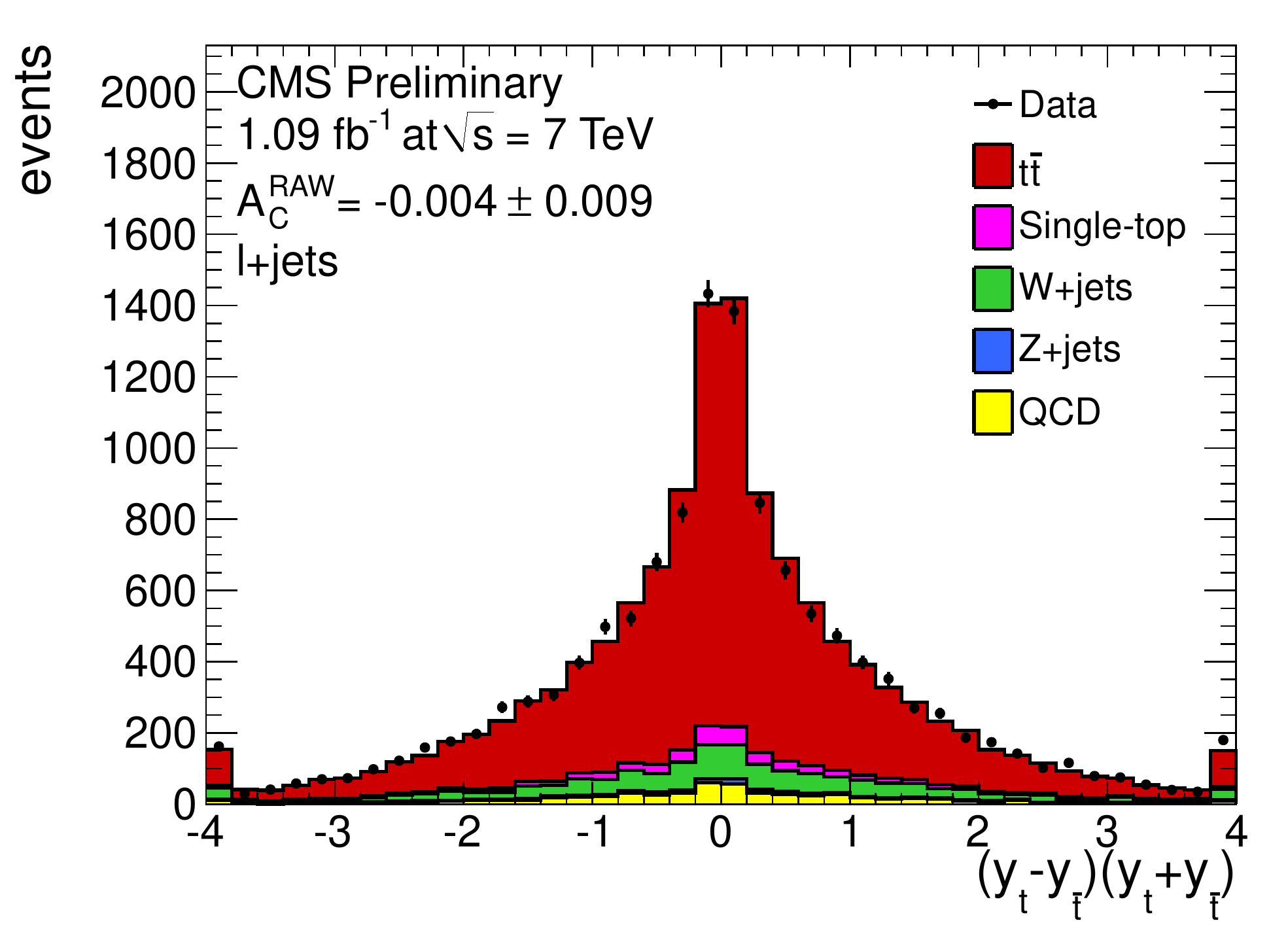}
\caption{Raw charge asymmetry in lepton+jet events.} \label{fig:Ac_raw}
\end{figure}

\begin{figure}[ht]
\centering
\includegraphics[width=0.48\textwidth]{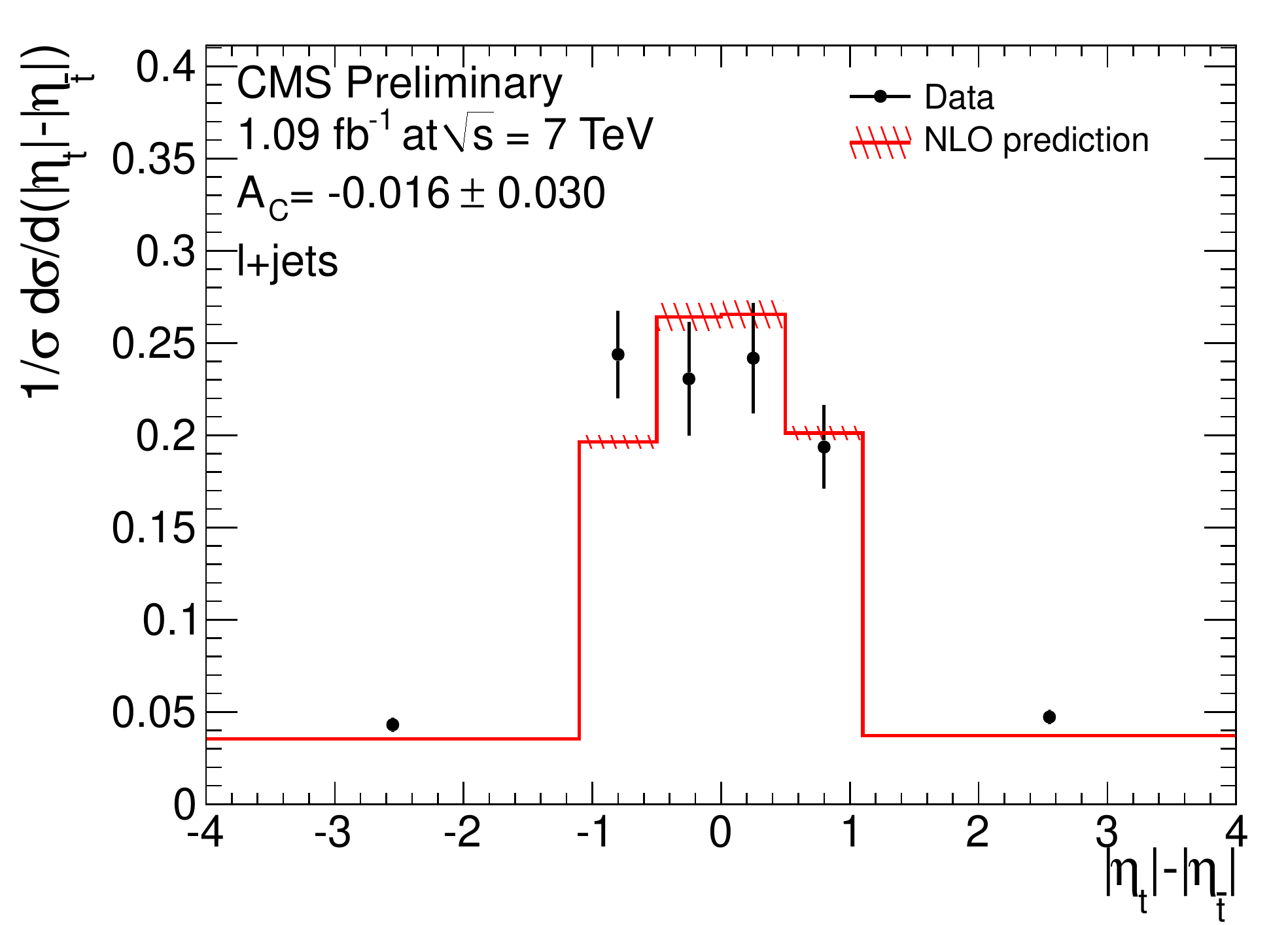}
\includegraphics[width=0.48\textwidth]{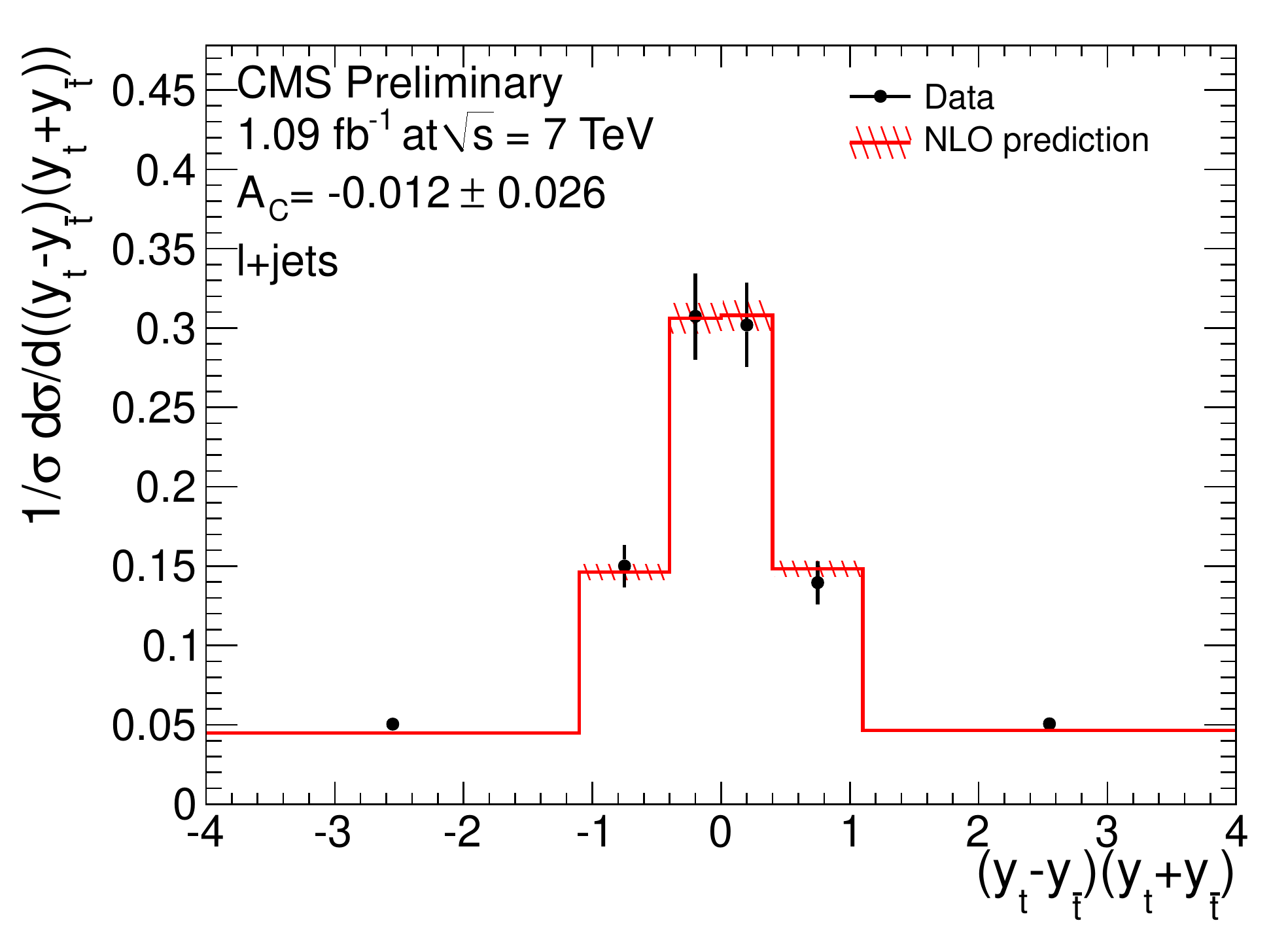}
\caption{Unfolded charge asymmetry in lepton+jet events.} \label{fig:Ac}
\end{figure}

Using 1.09 \invfb, CMS has measured the charge asymmetry in a clean
lepton+jets sample, selected using b-jet tagging and kinematic cuts
\cite{top-11-014}.  A kinematic fit is used to pick the best jet
combination and the charge asymmetry is constructed using
the pseudorapidity $\eta$ or rapidity $y$ in two ways,
$A_c^{\eta}=|\eta_t| - |\eta_{\bar{t}}|$ and 
$A_c^y= (y_t^2-y_{\bar{t}}^2) = (y_t-y_{\bar{t}}) (y_t+y_{\bar{t}})$.  
The raw asymmetries are shown in Fig.~\ref{fig:Ac_raw}, illustrating
the purity of the selected \ttbar events.
After unfolding for acceptance and resolution, no statistically
significant asymmetry is observed in either variable (Fig.~\ref{fig:Ac}):
$A_c^\eta=-0.016\pm 0.030^{+0.010}_{-0.019}$ compared to a SM
expectation of $0.013\pm0.001$ or 
$A_c^y=-0.013\pm0.026^{+0.026}_{-0.021}$ compared to a SM expectation
of $0.011\pm 0.001$.  Additionally, no trend appears when plotted against the \ttbar
invariant mass.  M.~Segala presented further details of the charge
asymmetry results at this meeting \cite{Segala}.

\section{Summary}
CMS has begun to investigate top quark production and decay, finding
agreement with the Standard Model expectations for top-pair production
through QCD and single-top production in the $t$-channel electroweak
process.  For \ttbar production, the cross sections measured using dilepton and
lepton plus jets topologies are in good agreement with approximate
next-to-next-to-leading order calculations, indicating good
understanding of top production at the LHC.  Detailed studies of
single-top production require more data, but CMS has observed
evidence for single top production.                                   

The top quark mass has been measured in \ttbar samples using the
dilepton and lepton+jets channels, extending techniques established at
the Tevatron.  These techniques work well and show promise for future
precision measurements at CMS.  Large samples will also open
the possibilities for new techniques that are less sensitive to
jet-energy scale uncertainties, which are currently the leading
systematic uncertainty.

CMS has also used top samples to search for new physics
contributions.  The \ttbar invariant mass spectrum is examined for
narrow resonance production, setting limits on production of a narrow
$Z^\prime$ or Kaluza-Klein gluon.  A new technique to reconstruct
boosted top quarks from the decay of massive particles using jet
substructure was demonstrated, extending the limits to higher masses.
By measuring the top charge asymmetry in (pseudo)rapidity, CMS limits
beyond-the-standard-model contributions.  However, additional data are
needed to shed light on the excess asymmetry observed in $p\bar{p}$
production.

\begin{acknowledgments}
Support for travel to this meeting and preparation of this
presentation was provided by the US Department of Energy 
grant number DE-FG05-97ER41031.
\end{acknowledgments}

\bigskip 

\end{document}